\documentclass[preprint, amsmath, amssymb, aps, pra]{revtex4-2}
\allowdisplaybreaks[4]

\usepackage{hyperref}
\hypersetup{
    pdfstartview = FitH,
    pdfauthor = {樊兆兴},
}

\usepackage[UseMSWordMultipleLineSpacing, MSWordLineSpacingMultiple = 1.5]{zhlineskip}

\usepackage{textgreek}
\newcommand{\mathpi}{\text{\textpi}}
\newcommand{\mathgamma}{\text{\textgamma}}

\usepackage{tikz}
\usepackage[compat = 1.1.0]{tikz-feynhand}
\tikzset{
    baseline = -2.56 pt,
}
\usetikzlibrary{backgrounds}
\usepackage{enumitem}
\setlist{nosep}

\usepackage{xcolor}
\definecolor{gray}{RGB}{248, 248, 248}

\begin{document}

\title{A Gauge Model for Quasi-Dirac Neutrinos}
\author{Zhao--Xing Fan}
\email{fanzhaoxing@itp.ac.cn}
\affiliation{Institute of Theoretical Physics, Chinese Academy of Sciences, Beijing 100190, China}
\affiliation{School of Physical Sciences, University of Chinese Academy of Sciences, Beijing 100049, China}
\author{Chun Liu}
\email{liuc@mail.itp.ac.cn}
\affiliation{Institute of Theoretical Physics, Chinese Academy of Sciences, Beijing 100190, China}
\affiliation{School of Physical Sciences, University of Chinese Academy of Sciences, Beijing 100049, China}
\date{\today}
\setlength{\parindent}{2 em}

\renewcommand{\abstractname}{\textbf{Abstract}}
\begin{abstract}
    \setlength{\parindent}{2 em}
    In a model with $L_{\text{\textmu}} - L_{\text{\texttau}}$ Abelian gauge symmetry, anomaly-free chiral fermions, which are Standard Model singlets, are introduced as the origin of right-handed neutrinos.
    This $\mathsf{U} (1)$ symmetry keeps the right-handed neutrinos Majorana massless.
    Tiny nonvanishing Dirac masses of neutrinos are due to higher-dimensional operators with natural coupling constants.
    After gauge symmetry breaking, a quasi-Dirac neutrino scenario naturally appears, and realistic neutrino physics can be produced.
    Phenomenological and cosmological aspects of the model are discussed.
\end{abstract}

\maketitle
\newpage

\section{Introduction}

Neutrino physics is the most exciting field in particle physics; the first concrete phenomenon beyond the Standard Model (SM) is neutrino oscillations, which means that neutrinos are not absolutely massless.
Roughly their typical mass is around the order of $10^{- 2}$ eV.
There are many new experiments to be carried out aiming at revealing neutrino properties.
In addition to neutrino oscillation experiments for CP violation and the mass ordering \cite{oscillation:DUNE:2015lol, oscillation:Hyper-Kamiokande:2016srs, oscillation:Li:2013zyd}, experiments for the neutrino Majorana property are of great importance \cite{Majorana:Agostini:2017iyd, Majorana:KamLAND-Zen:2016pfg, Majorana:McDonald:2017izm, Majorana:Chen:2016qcd}.
Massive neutrinos are classified into two types, one is Majorana, the other is Dirac, according to whether the masses originate from lepton number violation.
If in the future when neutrinoless double \textbeta{} decay ($0 \nu 2${}\textbeta) experiments reach the precision level of $0.005$ eV, and there are still no positive signals, neutrinos should be of the Dirac type.
In this paper, we consider a gauge model for quasi-Dirac neutrinos.

Dirac neutrinos always involve in right-handed neutrinos, and the right-handed neutrinos need to be Majorana mass-free.
The lepton number is conserved absolutely.
Meanwhile, neutrino Majorana masses, which are supposed to be much smaller than neutrino Dirac masses, might still be allowed.
The latter is the so-called quasi-Dirac neutrino scenario \cite{quasi-Dirac:Valle:1982yw, quasi-Dirac:Anamiati:2017rxw}.
In this case, the lepton number is only conserved approximately.

In the Dirac or quasi-Dirac scenario, small masses usually imply very small coupling constants in theoretical models.
To avoid the naturalness problem, the so-called Dirac seesaw mechanism \cite{DiracSeesaw:Roncadelli:1983ty, DiracSeesaw:Babu:1988yq, DiracSeesaw:Ma:2016mwh} is needed.
This can be achieved by introducing some global flavor symmetries \cite{Dirac-neutrino:Reyimuaji:2024kqs, Chang:1999pb}.
Alternatively, gauge symmetries can also be considered.
For simplicity, we exploit a well-motivated Abelian gauge symmetry, which is known as the $L_{\text{\textmu}} - L_{\text{\texttau}}$ family symmetry \cite{He:1991qd}.

\section{The Model}

Besides the SM $\mathsf{SU} (3)_{\text{C}} \times \mathsf{SU} (2)_{\text{L}} \times \mathsf{U} (1)_{Y}$ gauge symmetry, a new $\mathsf{U} (1)$ gauge symmetry is introduced.
While quarks and the SM Higgs are neutral under the new $\mathsf{U} (1)$, it concerns the lepton sector only.
We denote the SM three generation lepton doublets as $L_{i} (z_{i})$ and lepton singlets as $e_{\text{R} i}^{\text{C}} (- z_{i})$, where $i = 1, 2, 3$ and $z_{i}$'s are new $\mathsf{U} (1)$ charges.
It is chosen as
\begin{equation}
    z_{1} = 0 \, , \qquad z_{2} = - z_{3} \, .  
\end{equation} 
While the first family is trivial, the second and third family SM leptons have opposite charges and thus are vector-like under this Abelian gauge symmetry dubbed as $\mathsf{U} (1)_{L_{\text{\textmu}} - L_{\text{\texttau}}}$.
There are no chiral anomalies.

For right-handed neutrinos, we further introduce six SM singlet Weyl fermions:
\begin{equation}
    \begin{alignedat}{3}
        & N_{1} (2 z) \, , & \qquad & N_{2} (2 z) \, , & \qquad & N_{3} (2 z) \, , \\
        & N_{4} (- 8 z) \, , & & N_{5} (- 8 z) \, , & & N_{6} (10 \, z) \, .
    \end{alignedat}
\end{equation}
Under such charge assignment, the chiral anomaly cancels \cite{anomaly-cancel:Geng:1989tcu, anomaly-cancel:He:1990me, anomaly-cancel:Kong:1996as, anomaly-cancel:Costa:2019zzy, Liu:2021ktw, anomaly-cancel:deGouvea:2015pea, anomaly-cancel:Wong:2020obo, anomaly-cancel:He:2025uvm}.
Actually, the number of chiral fields which are SM singlets is at least five, and the charge assignment is not unique.
For our purpose, we choose six chiral fermions with above $\mathsf{U} (1)_{L_{\text{\textmu}} - L_{\text{\texttau}}}$ charges.
In addition, four dark Brout--Englert--Higgs fields for $\mathsf{U} (1)_{L_{\text{\textmu}} - L_{\text{\texttau}}}$ breaking are chosen as 
\begin{equation}
    \varphi_{1} (- z) \, , \qquad \varphi_{2} (- 2 z) \, , \qquad \varphi_{3} (- 3 z) \, , \qquad \varphi_{4} (16 \, z) \, .  
\end{equation}

It should be first noted that, under the $\mathsf{U} (1)_{L_{\text{\textmu}} - L_{\text{\texttau}}}$ charge assignment, in the Lagrangian, Yukawa coupling terms of the SM Higgs field $H$ and charged leptons are flavor diagonal automatically.
Weak eigenstates of charged leptons are identical to their own mass eigenstates.
\begin{equation}
    \mathcal{L} \supset - \sqrt{2} \bigl( y_{1 1}^{e} H^{\dagger} L_{1} e_{\text{R} 1}^{\text{C}} + y_{2 2}^{e} H^{\dagger} L_{2} e_{\text{R} 2}^{\text{C}} + y_{3 3}^{e} H^{\dagger} L_{3} e_{\text{R} 3}^{\text{C}} + \text{h.c.} \bigr) \, ,
\end{equation}
where $y^{e}$'s denote Yukawa couplings, and fermions are written in terms of 2-component Weyl spinors.

By choosing $z_{2} = z$, the renormalizable gauge invariant Lagrangian involving above $\mathsf{U} (1)_{L_{\text{\textmu}} - L_{\text{\texttau}}}$ relevant fields is following.
\begin{equation}    
\begin{aligned}
    \mathcal{L}^{d \leqslant 4} & = - \frac{1}{4} \tilde{X}_{\mu \nu} \tilde{X}^{\mu \nu} + \frac{1}{2} \varepsilon B_{\mu \nu} \tilde{X}^{\mu \nu} \\
        & \mathrel{\phantom{=}} {} + \sum_{1 \leqslant i \leqslant 3} \bigl( \mathrm{i} \bar{L}_{i} \bar{\sigma}^{\mu} \, \mathrm{D}_{\mu} L_{i} + \mathrm{i} \bar{e}_{\text{R} i}^{\text{C}} \bar{\sigma}^{\mu} \, \mathrm{D}_{\mu} e_{\text{R} i}^{\text{C}} + \mathrm{i} \bar{N}_{i} \bar{\sigma}^{\mu} \, \mathrm{D}_{\mu} N_{i} + \mathrm{i} \bar{N}_{i + 3} \bar{\sigma}^{\mu} \, \mathrm{D}_{\mu} N_{i + 3} \bigr) \\
        & \mathrel{\phantom{=}} {} + \sum_{1 \leqslant k \leqslant 4} (\mathrm{D}_{\mu} \varphi_{k})^{\ast} \, \mathrm{D}^{\mu} \varphi_{k} + \mu_{k}^{2} \varphi_{k}^{\ast} \varphi_{k} - 2 \lambda_{H k} \varphi_{k}^{\ast} \varphi_{k} H^{\dagger} H \bigr) \\
        & \mathrel{\phantom{=}} {} - \sum_{1 \leqslant k, \ell \leqslant 4} \lambda_{k \ell} \varphi_{k}^{\ast} \varphi_{k} \varphi_{\ell}^{\ast} \varphi_{\ell} \\
        & \mathrel{\phantom{=}} {} + \bigl( \mu_{1 1 2} \varphi_{1}^{2} \varphi_{2}^{\ast} + \mu_{1 2 3} \varphi_{1} \varphi_{2} \varphi_{3}^{\ast} - \lambda_{1 1 1 3} \varphi_{1}^{3} \varphi_{3}^{\ast} - \lambda_{1 2 2 3} \varphi_{1}^{\ast} \varphi_{2}^{2} \varphi_{3}^{\ast} + \text{h.c.} \bigr) \\
        & \mathrel{\phantom{=}} {} - \frac{1}{\sqrt{2}} \Bigl( \sum_{4 \leqslant \alpha, \beta \leqslant 5} y_{\alpha \beta}^{N} \varphi_{4} N_{\alpha} N_{\beta} + 2 \sum_{4 \leqslant \alpha \leqslant 5} y_{\alpha 6}^{N} \varphi_{2} N_{\alpha} N_{6} + \text{h.c.} \Bigr) \, ,
\end{aligned} \label{eq: dim 4}
\end{equation}
where covariant derivatives $\mathrm{D}_{\mu}$'s contain appropriate gauge potentials $A_{\mu}^{a}$, $B_{\mu}$ and $\tilde{X}_{\mu}$ with gauge couplings $g_{2}$, $g_{1}$ and $g_{\text{N}}$ for $\mathsf{SU} (2)_{\text{L}}$, $\mathsf{U} (1)_{Y}$ and $\mathsf{U} (1)_{L_{\text{\textmu}} - L_{\text{\texttau}}}$, respectively.
$B_{\mu \nu}$ and $\tilde{X}_{\mu \nu}$ are the gauge field strengths of $\mathsf{U} (1)_{Y}$ and $\mathsf{U} (1)_{L_{\text{\textmu}} - L_{\text{\texttau}}}$, respectively.
$\varepsilon$ is the mixing parameter which is expected to be the magnitude of the one-loop correction.
$\mu$'s are mass parameters.
$\lambda$'s are dimensionless parameters.

The dimension $5$ Lagrangian is
\begin{equation}
\begin{aligned}
    \mathcal{L}^{d = 5} & = - \frac{2}{M} \sum_{1 \leqslant i \leqslant 3} \bigl( y_{1 i}^{\nu} \varphi_{2} \tilde{H}^{\dagger} L_{1} N_{i} + y_{2 i}^{\nu} \varphi_{3} \tilde{H}^{\dagger} L_{2} N_{i} + y_{3 i}^{\nu} \varphi_{1} \tilde{H}^{\dagger} L_{3} N_{i} \bigr) \\
        & \mathrel{\phantom{=}} {} - \frac{1}{M} \sum_{1 \leqslant i, j \leqslant 3} \bigl( y_{i j}^{N} \varphi_{2}^{2} + y_{i j}^{N \prime} \varphi_{1} \varphi_{3} \bigr) N_{i} N_{j} - \frac{2}{M} \sum_{\substack{1 \leqslant i \leqslant 3 \\ 4 \leqslant \alpha \leqslant 5}} y_{i \alpha}^{N} (\varphi_{3}^{\ast})^{2} N_{i} N_{\alpha} \\
        & \mathrel{\phantom{=}} {} - \frac{2}{M} \sum_{4 \leqslant \alpha \leqslant 5} \bigl( y_{\alpha 6}^{N} \varphi_{1}^{2} + y_{\alpha 6}^{N \prime} \varphi_{1}^{\ast} \varphi_{3} \bigr) N_{\alpha} N_{6} + \text{h.c.} \, , \\
\end{aligned} \label{eq: dim 5}
\end{equation}
where $\tilde{H}$ is defined as $\mathrm{i} \sigma^{2} H^{\ast}$, $M$ is a high energy scale of the new physics expected to be $10^{14}$ GeV.

\subsection{Neutrinos}

In the Lagrangian, there are no trilinear Dirac neutrino mass terms due to the $\mathsf{U} (1)_{L_{\text{\textmu}} - L_{\text{\texttau}}}$ symmetry.
Dirac neutrino masses originate from quadrilinear terms which are $\dfrac{1}{M}$ suppressed.
This provides a natural explanation for small Yukawa couplings of neutrinos.
Furthermore, once the bilinear neutrino mass terms for the right-handed neutrinos $N_{1}$, $N_{2}$ and $N_{3}$ are absent, a natural Dirac neutrino scenario is realized.

Let us go into more details of the model.
It is assumed that the scalar potential of this model leads to spontaneous symmetry breaking (SSB) of $\mathsf{SU} (2)_{\text{L}} \times \mathsf{U} (1)_{Y} \times \mathsf{U} (1)_{L_{\text{\textmu}} - L_{\text{\texttau}}}$ into $\mathsf{U} (1)_{\text{EM}}$.
For our purpose, fields $H$, $\varphi_{1}$, $\varphi_{2}$, $\varphi_{3}$ and $\varphi_{4}$ have non-vanishing vacuum expectation values (VEVs) $v_{H}$, $v_{1}$, $v_{2}$, $v_{3}$ and $v_{4}$, respectively.
In the unitarity gauge, they are expanded around the vacuum.
\begin{equation}
\begin{aligned}
    & H = \frac{1}{\sqrt{2}} (v_{H} + h) {\begin{pmatrix}
        0 \\
        1
    \end{pmatrix}} \, , \\
    & \varphi_{k} = \frac{1}{\sqrt{2}} (v_{k} + \phi_{k}) \, , \qquad (1 \leqslant k \leqslant 4) \, .
\end{aligned}
\end{equation}
Look at the SSB of $\mathsf{U} (1)_{L_{\mathrm{\mu}} - L_{\mathrm{\tau}}}$.
We note that for the case of non-single scalar fields, their VEVs are not necessarily of the same order of magnitude.
Rather a hierarchical pattern is more interesting, such as $v_{1}, v_{2}, v_{3} \ll v_{4}$.
In this case, it is $v_{4}$ which determines the gauge boson mass, whereas the other $v_{i}$'s contribute to fermion masses.
Thus $v_{4}$ is taken to be several TeV.

After SSB, by denoting $L_{i} = {\begin{pmatrix} \nu_{i} \\ e_{\text{L} i} \end{pmatrix}}$, Dirac neutrino masses are
\begin{equation}
    \mathcal{L}_{\text{D}} = - {\begin{pmatrix}
        \nu_{1} & \nu_{2} & \nu_{3}
    \end{pmatrix}} \frac{v_{H}}{M} {\begin{pmatrix}
        y_{1 1}^{\nu} v_{2} & y_{1 2}^{\nu} v_{2} & y_{1 3}^{\nu} v_{2} \\[1 ex]
        y_{2 1}^{\nu} v_{3} & y_{2 2}^{\nu} v_{3} & y_{2 3}^{\nu} v_{3} \\[1 ex]
        y_{3 1}^{\nu} v_{1} & y_{3 2}^{\nu} v_{1} & y_{3 3}^{\nu} v_{1}
    \end{pmatrix}} {\begin{pmatrix}
        N_{1} \\[1 ex]
        N_{2} \\[1 ex]
        N_{3}
    \end{pmatrix}} + \text{h.c.} \, . \label{eq: Dirac}
\end{equation}
It seems that this Dirac mass matrix is general enough to easily reproduce the realistic neutrino physics.
However, because charged leptons are flavor-diagonal, the neutrino Dirac mass matrix itself fixes both neutrino masses and mixing.
It is quite constrained.
For example, if neutrino masses are hierarchical, we cannot attribute this hierarchy to that of $v_{i}$'s, because in this case, large neutrino mixing could not be achieved.
Instead, certain degeneracy of $v_{i}$'s and $y_{i j}^{\nu}$'s are needed.

The Dirac mass matrix can be diagonalized by two unitary matrices $U$ and $V$:
\begin{equation}    
\begin{aligned}
    \mathcal{L}_{\text{D}} & = - {\begin{pmatrix}
        \nu_{1}^{\text{m}} & \nu_{2}^{\text{m}} & \nu_{3}^{\text{m}}
    \end{pmatrix}} U^{\text{T}} \frac{v_{H}}{M} {\begin{pmatrix}
        y_{1 1}^{\nu} v_{2} & y_{1 2}^{\nu} v_{2} & y_{1 3}^{\nu} v_{2} \\[1 ex]
        y_{2 1}^{\nu} v_{3} & y_{2 2}^{\nu} v_{3} & y_{2 3}^{\nu} v_{3} \\[1 ex]
        y_{3 1}^{\nu} v_{1} & y_{3 2}^{\nu} v_{1} & y_{3 3}^{\nu} v_{1}
    \end{pmatrix}} V {\begin{pmatrix}
        N_{1}^{\text{m}} \\[1 ex]
        N_{2}^{\text{m}} \\[1 ex]
        N_{3}^{\text{m}}
    \end{pmatrix}} + \text{h.c.} \\[1 ex]
    & = - {\begin{pmatrix}
        \nu_{1}^{\text{m}} & \nu_{2}^{\text{m}} & \nu_{3}^{\text{m}}
    \end{pmatrix}} {\begin{pmatrix}
        m_{1}^{\nu} \\[1 ex]
        & m_{2}^{\nu} \\[1 ex]
        & & m_{3}^{\nu}
    \end{pmatrix}} {\begin{pmatrix}
        N_{1}^{\text{m}} \\[1 ex]
        N_{2}^{\text{m}} \\[1 ex]
        N_{3}^{\text{m}}
    \end{pmatrix}} + \text{h.c.} \, ,
\end{aligned} \label{eq: Dirac diag}
\end{equation}
where the superscript m stands for mass eigenstates, and $\displaystyle \max_{1 \leqslant i \leqslant 3} m_{i}^{\nu} \sim \max_{1 \leqslant i \leqslant 3} \dfrac{y^{\nu}}{M} v_{i} v_{H}$.
$U$ is just the Pontecorvo--Maki--Nakagawa--Sakata (PMNS) matrix.
To achieve neutrino masses to be about $10^{- 3} \text{--} 10^{- 2}$ eV, $\displaystyle \max_{1 \leqslant i \leqslant 3} \dfrac{y^{\nu} v_{i}}{M}$ should be taken as $10^{- 14} \text{--} 10^{- 13}$.

Neutrino mixing can be seen in the following way.
It is noted that
\begin{equation}
    U^{\ast} {\begin{pmatrix}
        (m_{1}^{\nu})^{2} \\[1 ex]
        & (m_{2}^{\nu})^{2} \\[1 ex]
        & & (m_{3}^{\nu})^{2}
    \end{pmatrix}} U^{\text{T}} = \frac{v_{H}^{2}}{M^{2}} {\begin{pmatrix}
        v_{2}^{2} \boldsymbol{y}_{1}^{\nu} \boldsymbol{y}_{1}^{\nu \dagger} & v_{2} v_{3} \boldsymbol{y}_{1}^{\nu} \boldsymbol{y}_{2}^{\nu \dagger} & v_{1} v_{2} \boldsymbol{y}_{1}^{\nu} \boldsymbol{y}_{3}^{\nu \dagger} \\[1 ex]
        v_{2} v_{3} \boldsymbol{y}_{2}^{\nu} \boldsymbol{y}_{1}^{\nu \dagger} & v_{3}^{2} \boldsymbol{y}_{2}^{\nu} \boldsymbol{y}_{2}^{\nu \dagger} & v_{1} v_{3} \boldsymbol{y}_{2}^{\nu} \boldsymbol{y}_{3}^{\nu \dagger} \\[1 ex]
        v_{1} v_{2} \boldsymbol{y}_{3}^{\nu} \boldsymbol{y}_{1}^{\nu \dagger} & v_{1} v_{3} \boldsymbol{y}_{3}^{\nu} \boldsymbol{y}_{2}^{\nu \dagger} & v_{1}^{2} \boldsymbol{y}_{3}^{\nu} \boldsymbol{y}_{3}^{\nu \dagger}
    \end{pmatrix}} \, , \label{eq: UMMU}
\end{equation}
where
\begin{equation}
    \boldsymbol{y}_{i}^{\nu} = {\begin{pmatrix}
        y_{i 1}^{\nu} & y_{i 2}^{\nu} & y_{i 3}^{\nu}
    \end{pmatrix}} \, .
\end{equation}
The left side of Eq.~\eqref{eq: UMMU} is expressed by physically measurable quantities.
Experimental data with standard parameterization of the PMNS matrix \cite{ParticleDataGroup:2024cfk} are the following,
\begin{equation}
\begin{aligned}
    & \Delta m_{2 1}^{2} \equiv (m_{2}^{\nu})^{2} - (m_{1}^{\nu})^{2} = 7.53 \times 10^{- 5} ~ \text{eV}^{2} \, , \\
    & \Delta m_{3 2}^{2} \equiv (m_{3}^{\nu})^{2} - (m_{2}^{\nu})^{2} = 2.455 \times 10^{- 3} ~ \text{eV}^{2} \, , \\
    & \theta_{2 3} = 48.33^{\circ} \, , \qquad \theta_{1 3} = 8.51^{\circ} \, , \qquad \theta_{1 2} = 33.65^{\circ} \, , \qquad \delta = 1.19 \, \mathrm{\pi} \, ,
\end{aligned}
\end{equation}
for neutrino mass normal ordering (N.O.), and
\begin{equation}
\begin{aligned}
    & \Delta m_{2 1}^{2} = 7.53 \times 10^{- 5} ~ \text{eV}^{2} \, , \qquad \Delta m_{3 2}^{2} = - 2.529 \times 10^{- 3} ~ \text{eV}^{2} \, , \\
    & \theta_{2 3} = 48.04^{\circ} \, , \qquad \theta_{1 3} = 8.51^{\circ} \, , \qquad \theta_{1 2} = 33.65^{\circ} \, , \qquad \delta = 1.19 \, \mathrm{\pi} \, ,
\end{aligned}
\end{equation}
for neutrino mass inverted ordering (I.O.).
For instance, if the lightest neutrino is massless, matrix elements of Eq.~\eqref{eq: UMMU} are the following,
\begin{alignat}{2}
    & {\begin{pmatrix}
        0.078 & - 0.204 - 0.154 \, \mathrm{i} & - 0.227 - 0.137 \, \mathrm{i} \\
        - 0.204 + 0.154 \, \mathrm{i} & 1.409 & 1.203 + 0.003 \, \mathrm{i} \\
        - 0.227 + 0.137 \, \mathrm{i} & 1.203 - 0.003 \, \mathrm{i} & 1.12
    \end{pmatrix}} \times 10^{- 3} ~ \text{eV}^{2} \, , & \qquad & (\text{N.O.}) \, , \\[1 ex]
    & {\begin{pmatrix}
        2.423 & 0.246 + 0.152 \, \mathrm{i} & 0.175 + 0.136 \, \mathrm{i} \\
        0.246 - 0.152 \, \mathrm{i} & 1.154 & - 1.219 + 0.003 \, \mathrm{i} \\
        0.175 - 0.136 \, \mathrm{i} & - 1.219 - 0.003 \, \mathrm{i} & 1.406
    \end{pmatrix}} \times 10^{- 3} ~ \text{eV}^{2} \, , & & (\text{I.O.}) \, .
\end{alignat}
The other example for N.O. is the $\nu_{1} \text{--} \nu_{2}$ near-degeneracy, namely $m_{1}^{\nu} = 0.020 ~ \text{eV}$, $m_{2}^{\nu} = 0.022 ~ \text{eV}$, and $m_{3}^{\nu} = 0.054 ~ \text{eV}$, matrix elements of Eq.~\eqref{eq: UMMU} are the following,
\begin{equation}
    {\begin{pmatrix}
        0.478 & - 0.204 - 0.154 \, \mathrm{i} & - 0.227 - 0.137 \, \mathrm{i} \\
        - 0.204 + 0.154 \, \mathrm{i} & 1.809 & 1.203 + 0.003 \, \mathrm{i} \\
        - 0.227 + 0.137 \, \mathrm{i} & 1.203 - 0.003 \, \mathrm{i} & 1.519
    \end{pmatrix}} \times 10^{- 3} ~ \text{eV}^{2} \, .
\end{equation}

We have noted that to get the maximal mixing angle $\theta_{2 3}$, in the right side matrix of Eq.~\eqref{eq: UMMU}, the second row (or column) and the third row (or column) should be equal, which is achieved by tuning Yukawa coupling vectors $\boldsymbol{y}_{2}^{\nu}$ and $\boldsymbol{y}_{3}^{\nu}$ to be nearly parallel, and taking that $v_{1} \simeq v_{3}$.
Similarly, the relatively large mixing angles $\theta_{1 2}$ and $\theta_{1 3}$ can be obtained by requiring $v_{2}$ to be a kind of smaller (or equal) than $v_{1}$ and $v_{3}$ for neutrino mass normal (or inverted) ordering, and taking $\boldsymbol{y}_{1}^{\nu}$ to be nearly orthogonal to both $\boldsymbol{y}_{2}^{\nu}$ and $\boldsymbol{y}_{3}^{\nu}$.
We see that in all the above cases, the Dirac mass matrix with the appropriate choice of parameters can indeed produce real neutrino physics.

There are Majorana mass terms among $N_{1, 2, 3}$ themselves, which are${} \sim y^{N} \dfrac{v_{2}^{2}}{M}$ and $y^{N} \dfrac{v_{1} v_{3}}{M}$.
We take that $v_{i} \ll v_{H}$, then these right-handed neutrino Majorana masses are much smaller than the Dirac masses.
To be specific, we assume that $v_{1}, v_{3} \sim 10$ GeV, and $v_{2} \sim 1 \text{--} 10$ GeV, these Majorana masses are about $10^{-5} \text{--} 10^{- 3}$ eV.
The Dirac nature of neutrinos will not significantly change.
In other words, because the right-handed neutrino Majorana masses are at least one order of magnitude smaller than the Dirac masses, there is an approximate lepton number conservation.

Weyl fermions $N_{4}, N_{5}, N_{6}$ also obtain non-vanishing Majorana masses.
After SSB, Majorana masses are
\begin{equation}
    \mathcal{L}_{\text{M}} = - \frac{1}{2} {\begin{pmatrix}
        N_{4} & N_{5} & N_{6}
    \end{pmatrix}} {\begin{pmatrix}
        y_{4 4}^{N} v_{4} & y_{4 5}^{N} v_{4} & y_{4 6}^{N} v_{2} \\[1 ex]
        y_{4 5}^{N} v_{4} & y_{5 5}^{N} v_{4} & y_{5 6}^{N} v_{2} \\[1 ex]
        y_{4 6}^{N} v_{2} & y_{5 6}^{2} v_{2} & 0
    \end{pmatrix}} {\begin{pmatrix}
        N_{4} \\[1 ex]
        N_{5} \\[1 ex]
        N_{6}
    \end{pmatrix}} + \text{h.c.} \, , \label{eq: mass of N456}
\end{equation}
where the dimension-$5$ terms are omitted because they are relatively too small.
The Majorana mass matrix can be diagonalized by one unitary matrix $P$.
It is easy to estimate eigenvalues of the above Majorana mass matrix, that is $m_{4}^{N} \sim m_{5}^{N} \sim y^{N} v_{4} \sim {}$TeV, while $m_{6}^{N} \sim y^{N} \dfrac{v_{2}^{2}}{v_{4}} \sim 0.1$ GeV.

In addition, one notes from Eq.~\eqref{eq: dim 5} that there are Majorana mass terms between $N_{1, 2, 3}$ and $N_{4, 5}$ which are about $y^{N} \dfrac{v_{3}^{2}}{M} \sim 10^{- 3}$ eV.
This mixing is further seesaw suppressed by $v_{4}$.
That is to say, this mixing is negligibly small.

More strictly, let us look at the complete neutrino mass matrix.
It is $9 \times 9$ one.
\begin{equation}
\begin{aligned}
    \mathcal{L}_{\text{D} + \text{M}} & = - \frac{1}{2} {\begin{pmatrix}
        \nu_{1} & \nu_{2} & \nu_{3} & N_{1} & N_{2} & N_{3} & N_{4} & N_{5} & N_{6}
    \end{pmatrix}} \\
        & \mathrel{\phantom{=}} \qquad {} \cdot {\begin{pmatrix}
            0 & \mathcal{D}_{3 \times 3} & 0 \\[1 ex]
            \mathcal{D}_{3 \times 3}^{\text{T}} & \mathcal{R}_{3 \times 3} & \mathcal{E}_{3 \times 3} \\[1 ex]
            0 & \mathcal{E}_{3 \times 3}^{\text{T}} & \mathcal{O}_{3 \times 3}
    \end{pmatrix}} \\
        & \mathrel{\phantom{=}} \qquad {} \cdot {\begin{pmatrix}
            \nu_{1} & \nu_{2} & \nu_{3} & N_{1} & N_{2} & N_{3} & N_{4} & N_{5} & N_{6}
        \end{pmatrix}}^{\text{T}} + \text{h.c.} \, ,
\end{aligned} \label{eq: 9 x 9}
\end{equation}
where sub-matrices $\mathcal{D}_{3 \times 3}$ is the Dirac mass matrix appeared in Eq.~\eqref{eq: Dirac},
\begin{equation}
    \mathcal{D}_{3 \times 3} = \frac{v_{H}}{M} {\begin{pmatrix}
        y_{1 1}^{\nu} v_{2} & y_{1 2}^{\nu} v_{2} & y_{1 3}^{\nu} v_{2} \\[1 ex]
        y_{2 1}^{\nu} v_{3} & y_{2 2}^{\nu} v_{3} & y_{2 3}^{\nu} v_{3} \\[1 ex]
        y_{3 1}^{\nu} v_{1} & y_{3 2}^{\nu} v_{1} & y_{3 3}^{\nu} v_{1}
    \end{pmatrix}} \sim 10^{- 2} ~ \text{eV} \, ,
\end{equation}
$\mathcal{R}_{3 \times 3}$ is the Majorana mass matrix of $N_{1, 2, 3}$ themselves,
\begin{equation}
    \mathcal{R}_{3 \times 3} = \frac{v_{2}^{2}}{M} {\begin{pmatrix}
        y_{1 1}^{N} & y_{1 2}^{N} & y_{1 3}^{N} \\[1 ex]
        y_{1 2}^{N} & y_{2 2}^{N} & y_{2 3}^{N} \\[1 ex]
        y_{1 3}^{N} & y_{2 3}^{N} & y_{3 3}^{N}
    \end{pmatrix}} + \frac{v_{1} v_{3}}{M} {\begin{pmatrix}
        y_{1 1}^{N \prime} & y_{1 2}^{N \prime} & y_{1 3}^{N \prime} \\[1 ex]
        y_{1 2}^{N \prime} & y_{2 2}^{N \prime} & y_{2 3}^{N \prime} \\[1 ex]
        y_{1 3}^{N \prime} & y_{2 3}^{N \prime} & y_{3 3}^{N \prime}
    \end{pmatrix}} \sim 10^{- 3} ~ \text{eV} \, , \label{eq: right-handed Majorana}
\end{equation}
$\mathcal{E}_{3 \times 3}$ is the Majorana mass matrix betweem $N_{1, 2, 3}$ and $N_{4, 5, 6}$,
\begin{equation}
    \mathcal{E}_{3 \times 3} = \frac{v_{3}^{2}}{M} {\begin{pmatrix}
        y_{1 4}^{N} & y_{1 5}^{N} & 0 \\[1 ex]
        y_{2 4}^{N} & y_{2 5}^{N} & 0 \\[1 ex]
        y_{3 4}^{N} & y_{3 5}^{N} & 0
    \end{pmatrix}} \sim 10^{- 3} ~ \text{eV} \, ,
\end{equation}
and $\mathcal{O}_{3 \times 3}$ appeared in Eq.~\eqref{eq: mass of N456},
\begin{equation}
    \mathcal{O}_{3 \times 3} = {\begin{pmatrix}
        y_{4 4}^{N} v_{4} & y_{4 5}^{N} v_{4} & y_{4 6}^{N} v_{2} \\[1 ex]
        y_{4 5}^{N} v_{4} & y_{5 5}^{N} v_{4} & y_{5 6}^{N} v_{2} \\[1 ex]
        y_{4 6}^{N} v_{2} & y_{5 6}^{2} v_{2} & 0
    \end{pmatrix}} \sim \text{TeV} \, . \label{eq: sub-matrix of N 456}
\end{equation}
If we just look at the $6 \times 6$ sub-matrix of the fields $(\nu_{1}, \nu_{2}, \nu_3, N_{1}, N_{2}, N_{3})$, the Dirac or quasi-Dirac property of active neutrinos is just what has been described through Eqs.~\eqref{eq: Dirac} and \eqref{eq: Dirac diag}.
The Dirac masses of active neutrinos are determined by $\mathcal{D}_{3 \times 3}$, they are perturbed by $\mathcal{R}_{3 \times 3}$.
The above analysis will be affected by $\mathcal{O}_{3 \times 3}$ of fields $(N_{4}, N_{5}, N_{6})$ only via a seesaw suppressed effect${} \sim \dfrac{\mathcal{E}_{3 \times 3}^{2}}{\mathcal{O}_{3 \times 3}} \sim 10^{- 18}$ eV which is negligible compared to $\mathcal{R}_{3 \times 3}$.
Therefore, our quasi-Dirac scenario is indeed justified.

\subsection{Gauge Interactions}

Introducing $\mathsf{U} (1)_{L_{\text{\textmu}} - L_{\text{\texttau}}}$ brings us new features in the interaction.
The mixing term of gauge field kinetic energy $\dfrac{1}{2} \varepsilon B_{\mu \nu} \tilde{X}^{\mu \nu}$ can be eliminated through the following field and coupling redefinition:
\begin{equation}
    \begin{aligned}
        & B_{\mu}' = B_{\mu} + \frac{\varepsilon}{\sqrt{1 - \varepsilon^{2}}} \tilde{X}_{\mu} \, , \\
        & \tilde{X}_{\mu}' = \frac{1}{\sqrt{1 - \varepsilon^{2}}} \tilde{X}_{\mu} \, , \\
        & g_{\text{N}}' = \sqrt{1 - \varepsilon^{2}} g_{\text{N}} \, .
    \end{aligned}
\end{equation}
Hereafter, the superscript $\prime$'s are omitted for convenience.
The covariant derivatives of the SM fields are then written in the following:
\begin{align*}
    & \mathrm{D}_{\mu} L_{1} = \biggl( \partial_{\mu} - \mathrm{i} g_{2} A_{\mu}^{a} T^{a} + \frac{1}{2} \mathrm{i} g_{1} B_{\mu} + \frac{1}{2} \mathrm{i} g_{\text{N}} \theta \tilde{X}_{\mu} \biggr) L_{1} \, , \\
    & \mathrm{D}_{\mu} L_{2} = \biggl( \partial_{\mu} - \mathrm{i} g_{2} A_{\mu}^{a} T^{a} + \frac{1}{2} \mathrm{i} g_{1} B_{\mu} - \mathrm{i} g_{\text{N}} \biggl( z - \frac{1}{2} \theta \biggr) \tilde{X}_{\mu} \biggr) L_{2} \, , \\
    & \mathrm{D}_{\mu} L_{3} = \biggl( \partial_{\mu} - \mathrm{i} g_{2} A_{\mu}^{a} T^{a} + \frac{1}{2} \mathrm{i} g_{1} B_{\mu} + \mathrm{i} g_{\text{N}} \biggl( z + \frac{1}{2} \theta \biggr) \tilde{X}_{\mu} \biggr) L_{3} \, , \\
    & \mathrm{D}_{\mu} e_{\text{R} 1}^{\text{C}} = \biggl( \partial_{\mu} + \mathrm{i} g_{1} B_{\mu} + \mathrm{i} g_{\text{N}} \theta \tilde{X}_{\mu} \biggr) e_{\text{R} 1}^{\text{C}} \, , \\
    & \mathrm{D}_{\mu} e_{\text{R} 2}^{\text{C}} = \biggl( \partial_{\mu} + \mathrm{i} g_{1} B_{\mu} - \mathrm{i} g_{\text{N}} (z - \theta) \tilde{X}_{\mu} \biggr) e_{\text{R} 2}^{\text{C}} \, , \\
    & \mathrm{D}_{\mu} e_{\text{R} 3}^{\text{C}} = \biggl( \partial_{\mu} + \mathrm{i} g_{1} B_{\mu} + \mathrm{i} g_{\text{N}} (z + \theta) \tilde{X}_{\mu} \biggr) e_{\text{R} 3}^{\text{C}} \, , \tag{\refstepcounter{equation}\theequation} \\
    & \mathrm{D}_{\mu} H = \biggl( \partial_{\mu} - \mathrm{i} g_{2} A_{\mu}^{a} T^{a} - \frac{1}{2} \mathrm{i} g_{1} B_{\mu} - \frac{1}{2} \mathrm{i} g_{\text{N}} \theta \tilde{X}_{\mu} \biggr) H \, , \\
    & \mathrm{D}_{\mu} Q_{k} = \biggl( \partial_{\mu} - \mathrm{i} g_{2} A_{\mu}^{a} T^{a} - \frac{1}{6} \mathrm{i} g_{1} B_{\mu} - \frac{1}{6} \mathrm{i} g_{\text{N}} \theta \tilde{X}_{\mu} \biggr) Q_{k} \, , \\
    & \mathrm{D}_{\mu} u_{k} = \biggl( \partial_{\mu} - \frac{2}{3} \mathrm{i} g_{1} B_{\mu} - \frac{2}{3} \mathrm{i} g_{\text{N}} \theta \tilde{X}_{\mu} \biggr) u_{k} \, , \\
    & \mathrm{D}_{\mu} d_{k} = \biggl( \partial_{\mu} + \frac{1}{3} \mathrm{i} g_{1} B_{\mu} - \frac{1}{3} \mathrm{i} g_{\text{N}} \theta \tilde{X}_{\mu} \biggr) d_{k} \, .
\end{align*}
where $\theta = \dfrac{\varepsilon}{\sqrt{1 - \varepsilon^{2}}} \dfrac{g_{1}}{g_{\text{N}}}$.
For completeness, quark cases are listed.
It is understood that a field with $\mathsf{U} (1)_{Y}$ hypercharge $Y$ additionally acquires a $\mathsf{U} (1)_{L_{\text{\textmu}} - L_{\text{\texttau}}}$ charge $\theta Y$ due to the mixing.

After SSB, while like in SM, one redefines gauge fields in the following,
\begin{equation}
\begin{aligned}
    & \mathrm{W}_{\mu}^{+} = \frac{1}{\sqrt{2}} \bigl( A_{\mu}^{1} - \mathrm{i} A_{\mu}^{2} \bigr) \, , \\
    & \mathrm{W}_{\mu}^{-} = \frac{1}{\sqrt{2}} \bigl( A_{\mu}^{1} + \mathrm{i} A_{\mu}^{2} \bigr) \, , \\
    & \tilde{Z}_{\mu} = \cos \theta_{\text{W}} \, A_{\mu}^{3} - \sin \theta_{\text{W}} \, B_{\mu} \, , \\
    & A_{\mu} = \sin \theta_{\text{W}} \, A_{\mu}^{3} + \cos \theta_{\text{W}} \, B_{\mu}
\end{aligned}
\end{equation}
with $\theta_{\text{W}}$ being the Weinberg angle, $\theta_{\text{W}} = \arctan \dfrac{g_{1}}{g_{2}}$.
The neutral gauge boson masses are seen from the Lagrangian,
\begin{equation}
    \mathcal{L}^{\text{GB}} \supset \frac{1}{2} M_{1}^{2} \biggl( \tilde{Z}_{\mu} - \frac{\varepsilon \sin \theta_{\text{W}}}{\sqrt{1 - \varepsilon^{2}}} \tilde{X}_{\mu} \biggr) \biggl( \tilde{Z}^{\mu} - \frac{\varepsilon \sin \theta_{\text{W}}}{\sqrt{1 - \varepsilon^{2}}} \tilde{X}^{\mu} \biggr) + \frac{1}{2} M_{2}^{2} \tilde{X}_{\mu} \tilde{X}^{\mu} \, ,
\end{equation}
where
\begin{equation}
\begin{aligned}
    & M_{1} = \frac{g_{2} v_{H}}{2 \cos\theta_{\text{W}}} \, , \\
    & M_{2} = g_{\text{N}} \bigl( z_{\varphi_{1}}^{2} v_{1}^{2} + z_{\varphi_{2}}^{2} v_{2}^{2} + z_{\varphi_{3}}^{2} v_{3}^{2} + z_{\varphi_{4}}^{2} v_{4}^{2} \bigr)^{1 / 2} \, .
\end{aligned}
\end{equation}
The photon field $A_{\mu}$ is massless like in SM, while $\tilde{Z}_{\mu}$ and $\tilde{X}_{\mu}$ have mass mixing.
The mass eigenstates are the following,
\begin{equation}
\begin{aligned}
    & \mathrm{Z}_{\mu} = \cos \alpha \, \tilde{Z}_{\mu} + \sin \alpha \, \tilde{X}_{\mu} \, , \\
    & X_{\mu} = - \sin \alpha \, \tilde{Z}_{\mu} + \cos \alpha \, \tilde{X}_{\mu} \, ,
\end{aligned}
\end{equation}
where $\alpha$ is the mixing angle,
\begin{equation}
    \alpha = \frac{1}{2} \arctan \frac{\dfrac{2 \varepsilon}{\sqrt{1 - \varepsilon^{2}}} \sin \theta_{\text{W}}}{\biggl( \dfrac{M_{2}}{M_{1}} \biggr)^{2} - 1 + \dfrac{\varepsilon^{2}}{1 - \varepsilon^{2}} \sin^{2} \theta_{\text{W}}} \, .
\end{equation}
The physical $\mathrm{Z}$-boson and $X$-boson mass eigenvalues are thus obtained,
\begin{equation}
    m_{\mathrm{Z}}^{2} = \frac{1}{2} \bigl( T + \sqrt{T^{2} - 4 D} \bigr) \, , \qquad m_{X}^{2} = \frac{1}{2} \bigl( T - \sqrt{T^{2} - 4 D} \bigr) \, ,
\end{equation}
where
\begin{equation}
    T = M_{1}^{2} \biggl( 1 + \frac{\varepsilon^{2}}{1 - \varepsilon^{2}} \sin^{2} \theta_{\text{W}} \biggr) + M_{2}^{2}, \qquad D = M_{1}^{2} M_{2}^{2}.
\end{equation}
Note that the $\mathrm{W}$-boson mass in our model is still $M_{1} \cos \theta_{\text{W}}$.

\subsection{Landau Pole}

The position of the Landau Pole of the new $\mathsf{U} (1)$ gauge interaction should be calculated, because too many matter fields may make the position problematic.
Analogous to QED, the $X_{\mu}$ wave function renormalization constant is
\begin{equation}
    Z_{X} = 1 - \frac{g_{\text{N}}^{2}}{(4 \mathpi)^{2}} \biggl( \frac{4}{3} \sum_{f} z_{f}^{2} + \frac{2}{3} \sum_{b} z_{b}^{2} \biggr) \frac{1}{\varepsilon} = 1 - \frac{8}{3} \digamma \frac{(z g_{\text{N}})^{2}}{(4 \mathpi)^{2}} \frac{1}{\varepsilon} \, ,
\end{equation}
where indices $f$ and $b$ refer to 2-component Weyl spinors and complex scalars, respectively.
For any Fermion $\psi$ with $\mathsf{U} (1)_{L_{\text{\textmu}} - L_{\text{\texttau}}}$ charge $z_{\psi}$, the wave function renormalization constant of $\psi$ is
\begin{equation}
    Z_{\psi} = 1 - 2 z_{\psi}^{2} \frac{g_{\text{N}}^{2}}{(4 \mathpi)^{2}} \frac{1}{\varepsilon} + \dotsb \, .
\end{equation}
The $g_{\text{N}}$ renormalization constant $Z_{\text{N}}$ satisfies
\begin{equation}
    Z_{\psi} Z_{\text{N}} \sqrt{Z_{X}} = 1 - 2 z_{\psi}^{2} \frac{g_{\text{N}}^{2}}{(4 \mathpi)^{2}} \frac{1}{\varepsilon} + \dotsb \, ,
\end{equation}
then it shows that
\begin{equation}
    Z_{\text{N}} = 1 + \frac{4}{3} \frac{(z g_{\text{N}})^{2}}{(4 \mathpi)^{2}} \frac{1}{\varepsilon} + \dotsb \, .
\end{equation}
The $\beta$-functions of $z g_{\text{N}}$ at one-loop level is as follows:
\begin{equation}
    \beta (z g_{\text{N}}) = \mu \frac{\mathrm{d}}{\mathrm{d} \mu} (z g_{\text{N}}) = \frac{(z g_{\text{N}})^{3}}{12 \, \mathpi^{2}} \digamma \, .
\end{equation}
The running of $\alpha_{\text{N}} \equiv \dfrac{1}{4 \mathpi} (z g_{\text{N}})^{2}$ is solved as
\begin{equation}
    \alpha_{\text{N}} = - \biggl( \frac{2 \digamma}{3 \mathpi} \Bigl( \ln \frac{\mu}{\varLambda_{\text{N}}} \Bigr) \biggr)^{-1} \, ,
\end{equation}
where $\varLambda_{\text{N}}$ is the Landau pole of our model:
\begin{equation}
    \varLambda_{\text{N}} = m_{X} \exp \bigg( \frac{3 \mathpi}{2 \digamma \alpha_{\text{N}}} \biggr) \, .
\end{equation}
In our model,
\begin{equation}
\begin{aligned}
    \digamma & = \frac{1}{2 z^{2}} \bigl( 2 z_{L_{2}}^{2} + z_{e_{2}}^{2} + 2 z_{L_{3}}^{2} + z_{e_{2}}^{2} + z_{N_{1}}^{2} + z_{N_{2}}^{2} + z_{N_{3}}^{2} + z_{N_{4}}^{2} + z_{N_{5}}^{2} + z_{N_{6}}^{2} \bigr) \\
        & \mathrel{\phantom{=}} {} + \frac{1}{4 z^{2}} \bigl( z_{\varphi_{1}}^{2} + z_{\varphi_{2}}^{2} + z_{\varphi_{3}}^{2} + z_{\varphi_{4}}^{2} \bigr) \\
    & = 190.5 \, .
\end{aligned}
\end{equation}
 
\section{Phenomenology Analysis}

The model parameters should be considered in detail.
For the mixing parameter $\varepsilon$, due to one-loop correction is about $10^{- 3}$, we expect it is no less than loop correction, so we take it as $\varepsilon \simeq \mathcal{O} {\left( 10^{- 2} \text{--} 10^{- 3} \right)}$.
The Large Hadron Collider (LHC) experiment has limited $m_{X}$ to be above $5$ TeV \cite{ATLAS:2019erb}, we will take $m_{X} \simeq (5 \text{--} 10) ~ \text{TeV}$.
Since $m_{X} \gg m_{\mathrm{Z}}$, $\tilde{Z}_{\mu}$--$\tilde{X}_{\mu}$ mass mixing $\alpha$ is much smaller than $\varepsilon$.
So we can take
\begin{equation}
    m_{\mathrm{Z}} = \frac{g_{1} v_{H}}{2 \cos\theta_{\text{W}}} \, , \qquad
        m_{X} = g_{\text{N}} \bigl( z_{\varphi_{1}}^{2} v_{1}^{2} + z_{\varphi_{2}}^{2} v_{2}^{2} + z_{\varphi_{3}}^{2} v_{3}^{2} + z_{\varphi_{4}}^{2} v_{4}^{2} \bigr)^{1 / 2} \, .
\end{equation}
Since $v_{4} \gg v_{1, 2, 3}$, $m_{X} \approx 16 \, z g_{\text{N}} v_{4}$
For the new gauge coupling $g_{\text{N}}$, we choose $g_{\text{N}} \sim \mathcal{O} (0.1 \text{--} 0.01)$ in order to make its value close to the SM ones, or to be consistent with possible gauge couplings unification.
Then $v_{4}$ should be approximately $10^{4} \text{--} 10^{5}$ GeV by taking $z$ is $\dfrac{1}{2}$.
This leads to that $\varLambda_{\text{N}} \sim 10^{57} ~ \text{GeV}$.
Therefore, this model does not have the Landau pole problem.

\subsection{New Four-Fermions Interaction}

Even if $\varepsilon$ is $10^{- 2}$, it can still be neglected in the analysis of neutrino physics.
From mass eigenstate fields $L_{i}$, $e_{\text{R} i}^{\text{C}}$ and $\nu_{i}^{\text{m}}$, $N_{i}^{\text{m}}$, ($i = 1, 2, 3$), 4-component Dirac spinors are introduced,
\begin{equation}
    \psi_{\mathrm{e}} = {\begin{pmatrix}
        e_{\text{L}_{1}} \\
        \bar{e}_{\text{R} 1}^{\text{C}}
    \end{pmatrix}} \, , \qquad \psi_{\text{\textmu}} = {\begin{pmatrix}
        e_{\text{L}_{2}} \\
        \bar{e}_{\text{R} 2}^{\text{C}}
    \end{pmatrix}} \, , \qquad \psi_{\text{\texttau}} = {\begin{pmatrix}
        e_{\text{L}_{3}} \\
        \bar{e}_{\text{R} 3}^{\text{C}}
    \end{pmatrix}} \, , \text{\makebox[3 em]{and}} \psi_{\nu_{i}} = {\begin{pmatrix}
        \nu_{i}^{\text{m}} \\
        \bar{N}_{i}^{\text{m}}
    \end{pmatrix}} \, .
\end{equation}
Their $\mathsf{U} (1)_{L_{\text{\textmu}} - L_{\tau}}$ gauge interaction and relevant mass terms are written as follows,
\begin{equation}
\begin{aligned}
    \mathcal{L} &\supset z g_{\text{N}} \bar{\psi}_{\text{\textmu}} \gamma^{\mu} \psi_{\text{\textmu}} X_{\mu} - z g_{\text{N}} \bar{\psi}_{\text{\texttau}} \gamma^{\mu} \psi_{\text{\texttau}} X_{\mu} \\
        &\mathrel{\phantom{\supset}} {} + \frac{1}{2} z g_{\text{N}} \sum_{1 \leqslant i, j \leqslant 3} O_{i j} \bar{\psi}_{\nu_{i}} \gamma^{\mu} \bigl( 1 - \gamma^{5} \bigr) \psi_{\nu_{j}} X_{\mu} \\
        &\mathrel{\phantom{\supset}} {} + z g_{\text{N}} \sum_{1 \leqslant k \leqslant 3} \bar{\psi}_{\nu_{k}} \gamma^{\mu} \bigl( 1 + \gamma^{5} \bigr) \psi_{\nu_{k}} X_{\mu} \\
        &\mathrel{\phantom{\supset}} {} - m_{\mathrm{e}} \bar{\psi}_{\mathrm{e}} \psi_{\mathrm{e}} - m_{\text{\textmu}} \bar{\psi}_{\text{\textmu}} \psi_{\text{\textmu}} - m_{\text{\texttau}} \bar{\psi}_{\text{\texttau}} \psi_{\text{\texttau}} - \sum_{1 \leqslant k \leqslant 3} m_{k}^{\nu} \bar{\psi}_{\nu_{k}} \psi_{\nu_{k}} \, ,
\end{aligned}
\end{equation}
where $O_{i j}$'s are elements of the matrix
\begin{equation}
    O \equiv U^{\dagger} {\begin{pmatrix}
        0 \\
        & 1 \\
        & & - 1
    \end{pmatrix}} U \, .
\end{equation}
Integrating out the heavy $X_{\mu}$ boson, following new 4-fermion interactions are obtained,
\begin{equation}
\begin{aligned}
    \mathcal{L}_{\text{eff}} & = - 4 \sqrt{2} \, G_{\text{N}} \bar{\psi}_{\text{\textmu}} \gamma^{\mu} \psi_{\text{\textmu}} \bar{\psi}_{\text{\textmu}} \gamma_{\mu} \psi_{\text{\textmu}} - 4 \sqrt{2} \, G_{\text{N}} \bar{\psi}_{\text{\texttau}} \gamma^{\mu} \psi_{\text{\texttau}} \bar{\psi}_{\text{\texttau}} \gamma_{\mu} \psi_{\text{\texttau}} \\
        & \mathrel{\phantom{=}} {} + 4 \sqrt{2} \, G_{\text{N}} \bar{\psi}_{\text{\textmu}} \gamma^{\mu} \psi_{\text{\textmu}} \bar{\psi}_{\text{\texttau}} \gamma_{\mu} \psi_{\text{\texttau}} \\
        & \mathrel{\phantom{=}} {} - 2 \sqrt{2} \, G_{\text{N}} \sum_{1 \leqslant i, j \leqslant 3} O_{i j} \bar{\psi}_{\text{\textmu}} \gamma^{\mu} \psi_{\text{\textmu}} \bar{\psi}_{\nu_{i}} \gamma_{\mu} \bigl( 1 - \gamma^{5} \bigr) \psi_{\nu_{j}} \\
        & \mathrel{\phantom{=}} {} + 2 \sqrt{2} \, G_{\text{N}} \sum_{1 \leqslant i, j \leqslant 3} O_{i j} \bar{\psi}_{\text{\texttau}} \gamma^{\mu} \psi_{\text{\texttau}} \bar{\psi}_{\nu_{i}} \gamma_{\mu} \bigl( 1 - \gamma^{5} \bigr) \psi_{\nu_{j}} \\
        & \mathrel{\phantom{=}} {} - \sqrt{2} \, G_{\text{N}} \sum_{1 \leqslant i, j, k, \ell \leqslant 3} O_{i j} O_{k \ell} \bar{\psi}_{\nu_{i}} \gamma^{\mu} \bigl( 1 - \gamma^{5} \bigr) \psi_{\nu_{j}} \bar{\psi}_{\nu_{k}} \gamma_{\mu} \bigl( 1 - \gamma^{5} \bigr) \psi_{\nu_{\ell}} \\
        & \mathrel{\phantom{=}} {} - 4 \sqrt{2} \, G_{\text{N}} \sum_{1 \leqslant k \leqslant 3} \bar{\psi}_{\text{\textmu}} \gamma^{\mu} \psi_{\text{\textmu}} \bar{\psi}_{\nu_{k}} \gamma_{\mu} \bigl( 1 + \gamma^{5} \bigr) \psi_{\nu_{k}} \\
        & \mathrel{\phantom{=}} {} + 4 \sqrt{2} \, G_{\text{N}} \sum_{1 \leqslant k \leqslant 3} \bar{\psi}_{\text{\texttau}} \gamma^{\mu} \psi_{\text{\texttau}} \bar{\psi}_{\nu_{k}} \gamma_{\mu} \bigl( 1 + \gamma^{5} \bigr) \psi_{\nu_{k}} \\
        & \mathrel{\phantom{=}} {} - 2 \sqrt{2} \, G_{\text{N}} \sum_{1 \leqslant i, j, k \leqslant 3} O_{i j} \bar{\psi}_{\nu_{i}} \gamma^{\mu} \bigl( 1 - \gamma^{5} \bigr) \psi_{\nu_{j}} \bar{\psi}_{\nu_{k}} \gamma_{\mu} \bigl( 1 + \gamma^{5} \bigr) \psi_{\nu_{k}} \\
        & \mathrel{\phantom{=}} {} - 4 \sqrt{2} \, G_{\text{N}} \sum_{1 \leqslant k, \ell \leqslant 3} \bar{\psi}_{\nu_{k}} \gamma^{\mu} \bigl( 1 + \gamma^{5} \bigr) \psi_{\nu_{k}} \bar{\psi}_{\nu_{\ell}} \gamma_{\mu} \bigl( 1 + \gamma^{5} \bigr) \psi_{\nu_{\ell}} \, ,
\end{aligned} \label{eq: 4-Fermi}
\end{equation}
among formulae, $G_{\text{N}} = \dfrac{\sqrt{2}}{8} \dfrac{(z g_{\text{N}})^{2}}{m_{X}^{2}}$ is defined by analogy with the Fermi constant $G_{\text{F}}$.

It is seen that the new physics at low energies mainly lies in the tauon and muon sectors, including their interactions with neutrinos.
Neutrinos can decay.
The heaviest neutrino $\psi_{\nu_{3}}$ can decay into three lighter ones, for example $\psi_{\nu_{3}} \to 2 \psi_{\nu_{1}} + \bar{\psi}_{\nu_{1}}$.
The corresponding Lagrangian is
\begin{equation}
    \mathcal{L} \supset - 2 \sqrt{2} \, G_{\text{N}} O_{1 3} \bar{\psi}_{\nu_{1}} \gamma^{\mu} \bigl( 1 - \gamma^{5} \bigr) \psi_{\nu_{3}} \bar{\psi}_{\nu_{1}} \gamma_{\mu} \bigl( (O_{1 1} + 1) - (O_{1 1} - 1) \gamma^{5} \bigr) \psi_{\nu_{1}} \, .
\end{equation}
Since $O_{1 1} \simeq - 0.17$, $O_{1 3} \simeq 0.5$, we estimate that the width of the aforementioned decay is
\begin{equation}
    \varGamma = \frac{G_{\text{N}}^{2} (m_{3}^{\nu})^{5}}{48 \, \mathpi^{3}} = \bigl( 6.6 \times 10^{52} ~ \text{s} \bigr)^{- 1}.
\end{equation}
The neutrinos can be regarded as stable with such a long lifetime.

\subsection{Collider Phenomenology}

Besides neutrino physics, this model shares the same phenomenology as other $\mathsf{U} (1)_{L_{\text{\textmu}} - L_{\text{\texttau}}}$ models with a heavy $X$ boson.
Let us briefly mention some points in the following.

\begin{enumerate}[label = \textbf{(\arabic*)}]

\item 
The mass ratio $\dfrac{m_{\mathrm{Z}}}{m_{\mathrm{W}}}$ will slightly deviate from the expectations of the SM,
\begin{equation}
    \Bigl( \frac{m_{\mathrm{Z}}}{m_{\mathrm{W}}} \Bigr)^{2} = \frac{1}{\cos^{2} \theta_{\text{W}}} \biggl( 1 - \frac{M_{1}^{2}}{M_{2}^{2} - M_{1}^{2}} \sin^{2} \theta_{\text{W}} \, \varepsilon^{2} \biggr) \, .
\end{equation}
Numerically, the deviation is about $10^{- 8}$ or even smaller.

\item 
For $\mathrm{Z}$ physics, we take into account the effect brought about by $\varepsilon$.
The decay width of the process $\mathrm{Z} \to \psi \bar{\psi}$ is
\begin{equation}
    \varGamma = \frac{m_{\mathrm{Z}}}{12 \, \mathpi} \biggl( 1 - 4 \Bigl( \frac{m_{\psi}}{m_{\mathrm{Z}}} \Bigr)^{2} \biggr)^{1 / 2} \biggl( \biggl( 1 + 2 \Bigl( \frac{m_{\psi}}{m_{\mathrm{Z}}} \Bigr)^{2} \biggr) V_{\psi}^{2} + \biggl( 1 - 6 \Bigl( \frac{m_{\psi}}{m_{\mathrm{Z}}} \Bigr)^{2} \biggr) A_{\psi}^{2} \biggr) \, ,
\end{equation}
where the parameters $V_{\psi}$ and $A_{\psi}$ are
\begin{equation}
\begin{alignedat}{3}
    & V_{\mathrm{e}}^{\text{N}} = V_{\mathrm{e}}^{\text{SM}} \, , & \qquad
        & V_{\text{\textmu}}^{\text{N}} = V_{\text{\textmu}}^{\text{SM}} + z g_{\text{N}} \alpha \, , & \qquad
        & V_{\text{\texttau}}^{\text{N}} = V_{\text{\texttau}}^{\text{SM}} - z g_{\text{N}} \alpha \, , \\
    & A_{\mathrm{e}}^{\text{N}} = A_{\mathrm{e}}^{\text{SM}} \, , &
        & A_{\text{\textmu}}^{\text{N}} = A_{\text{\textmu}}^{\text{SM}} \, , &
        & A_{\text{\texttau}}^{\text{N}} = A_{\text{\texttau}}^{\text{SM}} \, .
\end{alignedat}
\end{equation}
There is no deviation from the SM results because $\alpha$ is too small.

\item Our muon $g - 2$ prediction is
\begin{equation}
    \Delta a_{\text{\textmu}} = \frac{\alpha_{\text{EM}}}{3 \mathpi \cos^{2} \theta_{\text{W}}} \Bigl( \frac{m_{\text{\textmu}}}{m_{X}} \Bigr)^{2} \bigl( z_{\varphi_{1}}^{2} + z_{\varphi_{2}}^{2} + z_{\varphi_{3}}^{2} + z_{\varphi_{4}}^{2} \bigr) \simeq 9.34 \times 10^{- 11} \, ,
\end{equation}
by using the result of Ref.~\cite{Baek:2001kca}.
This value is far smaller than the current world average \cite{Muong-2:2024hpx}:
\begin{equation}
    a_{\text{\textmu}}^{\text{exp}} - a_{\text{\textmu}}^{\text{SM}} = 2.55 \times 10^{-9} \, .
\end{equation}

\item This model can be verified directly at muon colliders \cite{MuonCollider:Ankenbrandt:1999cta} with a center-of-mass energy $\sqrt{s} \geqslant m_{X} > 5 ~ \text{TeV}$.
The collision of $\text{\textmu}^{-} \text{\textmu}^{+}$ produces on-shell $X_{\mu}$ and a photon.
Then, $X_{\mu}$ may decay into $\text{\textmu}^{-} \text{\textmu}^{+}$, $\text{\texttau}^{-} \text{\texttau}^{+}$ and so on.
By measuring the energy of the photons, $m_{X}$ can be obtained.
The existence of $X_{\mu}$ can be further verified through measuring the energy and momentum of muons from the decay and calculating their invariant mass.

\end{enumerate}

\subsection{Cosmological Consideration}

There are three right-handed neutrino flavors which are light in quasi-Dirac neutrino models.
The successful Big Bang Nucleosynthesis (BBN) can constrain the number of d.o.f.~of light particles in the thermal equilibrium \cite{Luo:2020sho, Ma:2026tyk, Langacker:2008yv}.
It is important to check right-handed neutrino production in cosmological evolution.
The Yukawa interaction given in Eq.~\eqref{eq: dim 5} is $\dfrac{1}{M}$ suppressed.
This suppresses production of right-handed neutrinos, even at the energy scale of $\mathsf{U} (1)_{L_{\text{\textmu}} - L_{\text{\texttau}}}$ breaking, namely at about $(5 \text{--} 100)$ TeV.
Important right-handed neutrino production might be due to new gauge interactions, as described in the effective Lagrangian \eqref{eq: 4-Fermi}.
It was shown in Ref.~\cite{Luo:2020sho} that to avoid the cosmological constraints, the new gauge boson is required to be heavier than $43$ TeV.
Thus, we simply let $m_{X} \geqslant 43$ TeV, by making $v_{4}$ several tens TeV or so.

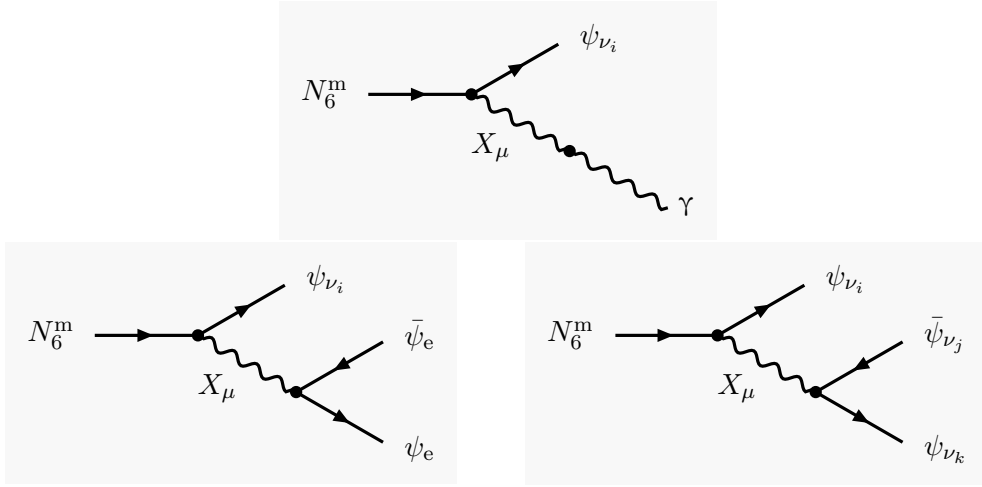
\begin{figure}[h]
\begin{tikzpicture}[framed, background rectangle/.style = {fill = gray}]
\begin{feynhand}
    \newcommand{\jiao}{30}
    \newcommand{\bi}{1.5}
    \vertex (N) at (- \bi, 0) {};
    \node[left] at (N) {$N_{6}^{\text{m}}$};
    \vertex[dot] (x) at (0, 0) {};
    \vertex[particle] (nu) at (\jiao: \bi) {};
    \node[right] at (nu) {$\psi_{\nu_{i}}$};
    \vertex[dot] (y) at (- \jiao: \bi) {};
    \begin{scope}[shift = (y)]
        \vertex (A) at (- \jiao: \bi);
        \node[right] at (A) {$\mathgamma$};
    \end{scope}
    \propag[fer] (N) to (x);
    \propag[fer] (x) to (nu);
    \propag[bos] (x) to[edge label' = $X_{\mu}$] (y);
    \propag[bos] (y) to (A);
\end{feynhand}
\end{tikzpicture} \\
\begin{tikzpicture}[framed, background rectangle/.style = {fill = gray}]
\begin{feynhand}
    \newcommand{\jiao}{30}
    \newcommand{\bi}{1.5}
    \vertex (N) at (- \bi, 0) {};
    \node[left] at (N) {$N_{6}^{\text{m}}$};
    \vertex[dot] (x) at (0, 0) {};
    \vertex[particle] (nu) at (\jiao: \bi) {};
    \node[right] at (nu) {$\psi_{\nu_{i}}$};
    \vertex[dot] (y) at (- \jiao: \bi) {};
    \begin{scope}[shift = (y)]
        \vertex[particle] (e +) at (\jiao: \bi) {};
        \node[right] at (e +) {$\bar{\psi}_{\mathrm{e}}$};
        \vertex[particle] (e -) at (- \jiao: \bi) {};
        \node[right] at (e -) {$\psi_{\mathrm{e}}$};
    \end{scope}
    \propag[fer] (N) to (x);
    \propag[bos] (x) to[edge label' = $X_{\mu}$] (y);
    \propag[fer] (x) to (nu);
    \propag[fer] (y) to (e -);
    \propag[antfer] (y) to (e +);
\end{feynhand}
\end{tikzpicture} \qquad
\begin{tikzpicture}[framed, background rectangle/.style = {fill = gray}]
\begin{feynhand}
    \newcommand{\jiao}{30}
    \newcommand{\bi}{1.5}
    \vertex[particle] (N) at (- \bi, 0) {};
    \node[left] at (N) {$N_{6}^{\text{m}}$};
    \vertex[dot] (x) at (0, 0) {};
    \vertex[particle] (nu) at (\jiao: \bi) {};
    \node[right] at (nu) {$\psi_{\nu_{i}}$};
    \vertex[dot] (y) at (- \jiao: \bi) {};
    \begin{scope}[shift = (y)]
        \vertex[particle] (e +) at (\jiao: \bi) {};
        \node[right] at (e +) {$\bar{\psi}_{\nu_{j}}$};
        \vertex[particle] (e -) at (- \jiao: \bi) {};
        \node[right] at (e -) {$\psi_{\nu_{k}}$};
    \end{scope}
    \propag[fer] (N) to (x);
    \propag[bos] (x) to[edge label' = $X_{\mu}$] (y);
    \propag[fer] (x) to (nu);
    \propag[fer] (y) to (e -);
    \propag[antfer] (y) to (e +);
\end{feynhand}
\end{tikzpicture}
\caption{The two-body decay and three-body decay of $N_{6}^{\text{m}}$.}
\label{fig: decay}
\end{figure}

The mass eigenstate corresponding to mass $m_{6}^{N}$, namly $N_{6}^{\text{m}}$, is a candidate for sub-GeV dark matter.
All decay modes of $N_{6}^{\text{m}}$ are shown in Fig.~\ref{fig: decay}.
The mixing between $N_{6}^{\text{m}}$ and $\psi_{\nu_{i}}$ is seesaw suppressed twice, namely mixing angle is that
\begin{equation}
    \vartheta = \frac{v_{2}}{v_{4}} \frac{\mathcal{R}_{3 \times 3}}{\mathcal{O}_{3 \times 3}} \sim 10^{- 17}.
\end{equation}
The decay widths are as follows,
\begin{equation}
\begin{aligned}
    & \varGamma \bigl( N_{6}^{\text{m}} \to \psi_{\nu_{i}} + \mathgamma \bigr) = \frac{1}{16 \, \mathpi} (z_{N_{6}} g_{\text{N}} \varepsilon \vartheta)^{2} m_{6}^{N} = \bigl( 3.3 \times 10^{18} ~ \text{s} \bigr)^{- 1}, \\
    & \varGamma \bigl( N_{6}^{\text{m}} \to \psi_{\nu_{i}} + \bar{\psi}_{\mathrm{e}} + \psi_{\mathrm{e}} \bigr) = \frac{1}{96 \, \mathpi^{3}} (z_{N_{6}} \varepsilon \vartheta G_{\text{N}})^{2} (m_{6}^{N})^{5} = \bigl( 3.9 \times 10^{42} ~ \text{s} \bigr)^{- 1}, \\
    & \varGamma \bigl( N_{6}^{\text{m}} \to \psi_{\nu_{i}} + \psi_{\nu_{j}} + \psi_{\nu_{k}} \bigr) \sim \varGamma \bigl( N_{6}^{\text{m}} \to \psi_{\nu_{i}} + \bar{\psi}_{\mathrm{e}} + \psi_{\mathrm{e}} \bigr).
\end{aligned}
\end{equation}
The lifetime of $N_{6}^{\text{m}}$ is far longer than the age of the universe.
The radioactive two-body decay is severely constrained unless the $N_{6}^{\text{m}}$ density in the universe is very small.
Nevertheless, this decay rate can be lowered via adjusting the $\varepsilon$ parameter.
If the radioactive decay is switched off, then the three-body decays can be important.
Extra electrons and positrons are expected in high-density places like that close to the galaxy center. \cite{Bouchet:2011fn}

\section{Summary and Discussion}

Neutrinos are Majorana or Dirac?
That is still a question nowadays.
In this work, we have tried to understand the Dirac neutrino scenario from a theoretical model.
A new $\mathsf{U} (1)$ gauge symmetry plays a key role in realizing such a scenario with no global or discrete symmetries introduced.
The right-handed neutrinos are taken as new chiral fermions which carry new $\mathsf{U} (1)$ charges.
Anomaly cancellation basically fixes the fermion content of the model.
Usually Dirac neutrino scenarios suffer from the naturalness problem compared to Majorana cases, but in our model, small neutrino masses are obtained in a similar way to the Froggatt--Nielsen mechanism which makes small neutrino masses as natural as the seesaw mechanism.

Actually, the model we have proposed is for quasi-Dirac neutrinos.
The gauge symmetry is taken as $\mathsf{U} (1)_{L_{\text{\textmu}} - L_{\text{\texttau}}}$.
Six SM singlet Weyl fermions are introduced for the $\mathsf{U} (1)_{L_{\text{\textmu}} - L_{\text{\texttau}}}$ anomaly cancellation, while three of them are right-handed neutrinos.
The Dirac masses are much larger than the Majorana masses.

Note that in the Lagrangian, we did not write down the Weinberg operator \cite{Weinberg:1979sa} which is usually considered as a result of integrating out the heavy right-handed neutrinos.
However, in this model, the right-handed neutrinos are very light.

This model can be directly verified at the future muon colliders, and the cosmological constraints can be satisfied whence the new $\mathsf{U} (1)_{L_{\text{\textmu}} - L_{\text{\texttau}}}$ physics scale is high enough.
Actually, our model parameters can be largely relaxed.
Our aim is to construct a model for quasi-Dirac neutrinos.

In the extreme case, we can take $v_{4}$ to be PeV scale.
Namely, the new physics scale will be PeV.
This results in less interesting phenomenology, but it may make the model cosmologically safer.

\begin{acknowledgments}
    This work was supported in part by the National Natural Science Foundation of China under the grant No.~12275335.
\end{acknowledgments}

\bibliography{Reference}

\end{document}